\documentclass{article}

\usepackage{PRIMEarxiv}

\usepackage[utf8]{inputenc} % allow utf-8 input
\usepackage[T1]{fontenc}    % use 8-bit T1 fonts
\usepackage{url}            % simple URL typesetting
\usepackage{booktabs}       % professional-quality tables
\usepackage{amsfonts}       % blackboard math symbols
\usepackage{nicefrac}       % compact symbols for 1/2, etc.
\usepackage{microtype}      % microtypography
\usepackage{lipsum}
\usepackage{fancyhdr}       % header
\usepackage{graphicx}       % graphics
\graphicspath{{media/}}     % organize your images and other figures under media/ folder

\usepackage{algorithm}
\usepackage{algorithmic}
\usepackage{amsfonts}
\usepackage{amsmath}
\usepackage{amssymb}
\usepackage{amsthm}
\usepackage{boxedminipage}
\usepackage{caption} 
\usepackage{enumitem}
\usepackage{graphicx}
\usepackage{multirow}
\usepackage{natbib}  
\usepackage{pifont}
\usepackage{pgfplots}
\usepackage{comment}
\usepackage{xcolor}         % colors
\usepackage{tikz}
\usepackage{thm-restate}
\setlist{nolistsep}
\usetikzlibrary{arrows}
\usetikzlibrary{patterns}

\newtheorem{definition}{Definition}%
\newtheorem{theorem}{Theorem}%
\newtheorem{example}{Example}

\DeclareMathOperator*{\argmax}{arg\,max}

\allowdisplaybreaks

\usepackage{authblk}

\title{\bf Achieving Balanced Representation in School Choice with Diversity Goals}
\author[1,2]{Zhaohong Sun}
\author[1]{Makoto Yokoo}

\affil[1]{Kyushu University, Japan}
\affil[2]{CyberAgent, Japan}

\date{\vspace{-10mm}}

\begin{document}

\maketitle

\begin{abstract}
Student placements under diversity constraints are a common practice globally. This paper addresses the selection of students by a single school under a \emph{one-to-one convention}, where students can belong to multiple types but are counted only once based on one type. While existing algorithms in economics and computer science aim to help schools meet diversity goals and priorities, we demonstrate that these methods can result in significant imbalances among students with different type combinations.

To address this issue, we introduce a new property called \emph{balanced representation}, which ensures fair representation across all types and type combinations. We propose a straightforward choice function that uniquely satisfies four fundamental properties: maximal diversity, non-wastefulness, justified envy-freeness, and balanced representation. While previous research has primarily focused on algorithms based on bipartite graphs, we take a different approach by utilizing flow networks. This method provides a more compact formalization of the problem and significantly improves computational efficiency. Additionally, we present efficient algorithms for implementing our choice function within both the bipartite graph and flow network frameworks.
\end{abstract}

\section{Introduction}

School choice programs with diversity considerations are widespread globally and have been extensively studied in the economics and computer science literature \citep{BFIM10a,EcYe15a,KHIY15a,GIKY+16a,KaKo18a,AGSW19a,BCC+19a}. In these programs, students are often associated with multiple types, such as gender, race, disability, or special talent, and schools typically establish quotas or targets for each type to achieve diverse and balanced outcomes.

Many studies implicitly adopt a \emph{one-to-all convention}, where each student is counted under every type they belong to, as observed in Israeli gap year programs \citep{GNKR19a}. In contrast, the \emph{one-to-one convention} is more practical in settings like college admissions in India \citep{SoYe22a} and school choice in Chile \citep{CEE+22a}. Under this convention, each student is counted under only one of their associated types. For example, an Aboriginal girl might occupy either a seat reserved for Aboriginal students or one reserved for girls, but not both. The one-to-one convention also applies in other critical contexts, such as vaccine allocation \citep{PSUY21a,AzBr21a}.
% In this setting, a centralized agent distributes vaccines to people who may belong to one or more priority categories, e.g. ages, disability, essential services. Once a person receives a vaccine, she counts only one of the categories for which she is qualified. 

The one-to-one convention has been extensively studied in the school choice literature, with most studies assuming that students have strict preferences over reserved seats of different types (or resolving ties to create a strict preference ordering) \citep{AyTu17a,KHIY17a,BCC+19a}. However, students are often indifferent to the type of reserved seat they are assigned, provided they secure a spot at their desired school. Imposing strict preferences over reserved types can lead to unintended consequences, potentially impacting the assignments of other students. Additionally, this approach does not always optimally fulfill diversity goals. To address these limitations, \citet{SoYe22a} proposed a novel method called ``smart reserves'' where all reserved seats aimed at achieving diversity goals are maximally satisfied without requiring students to express preferences over different types of reserved seats. \citet{AzSu21a} further extended this concept to a more general framework with ``multiple ranks of reserves,'' accommodating a broader spectrum of diversity constraints such as minimum and maximum quotas and proportionality requirements.

An important issue that has not been adequately addressed in previous work is the potential for imbalanced outcomes among students with distinct type combinations. For instance, consider a school that reserves seats for both aborigines and girls. Existing algorithms might lead to scenarios where the seats reserved for aborigines are filled exclusively by aboriginal boys, while the seats reserved for girls are occupied by non-aboriginal students. Consequently, no aboriginal girls are matched, even though the quotas for both types are technically met. Such an outcome fails to achieve the original objective of fostering a diverse and balanced integration of students from different backgrounds. Ideally, the goal is to compute an assignment that ensures fair representation not only for each individual type but also for combinations of types.

Another critical issue is computational efficiency. This concern becomes particularly relevant because, in practice, the number of privilege types is typically small, leading to a significantly smaller number of distinct type combinations compared to the number of students. For instance, in Brazilian college admissions, there are three privilege types: public high school, low-income, and racial minority \citep{AyBo20a}. Similarly, in Indian college admissions, the social privilege groups include Scheduled Castes (SCs), Scheduled Tribes (STs), and Other Backward Classes (OBCs), alongside additional reservation categories such as gender and disability \citep{SoYe22a}.

Existing selection rules for schools are based on bipartite graph structures that model the relationship between students and reserved seats \citep{SoYe22a, AzSu21a}. In contrast, we show that the problem can be represented more compactly by leveraging flow networks, where nodes correspond to types and type combinations. This approach offers a key advantage: the associated flow network problem can be solved in strongly polynomial time, making it independent of the number of students.

In this paper, we adopt the model of multiple ranks of quotas and follow the one-to-one matching convention. While maintaining ranked quotas for different types as the primary diversity goals, we aim to address imbalances among students with different type combinations to promote a more equitable and integrated outcome. We summarize our contributions as follows:

 First, we introduce a new property called \emph{balanced representation} to address the issue of imbalance. This property aims to maximize the minimum selection ratio across all matchings that adhere to maximal diversity.

Second, we propose a straightforward choice function for schools that \emph{uniquely} satisfies four key properties: i) maximal diversity, ii) non-wastefulness, iii) justified envy-freeness, and iv) balanced representation.

Third, we develop a novel graph structure based on flow networks, providing a more compact representation of the problem compared to existing methods.

Fourth, we present efficient algorithms to implement the newly proposed choice function using two graph structures: ranked reservation graphs, as studied in previous work, and flow networks, which we propose.

% \begin{itemize}
% \item First, we introduce a new property called \emph{balanced representation} to address the issue of imbalance. This property aims to maximize the minimum selection ratio across all matchings that adhere to maximal diversity.
% \item Second, we propose a straightforward choice function for schools that \emph{uniquely} satisfies four key properties: i) maximal diversity, ii) non-wastefulness, iii) justified envy-freeness, and iv) balanced representation.
% \item Third, we develop a novel graph structure based on flow networks, providing a more compact representation of the problem compared to existing methods.
% \item Fourth, we present efficient algorithms to implement the newly proposed choice function using two graph structures: ranked reservation graphs, as studied in previous work, and flow networks, which we propose.
% \end{itemize}

% Due to space limitations, a detailed review of recent developments in school choice with diversity goals is provided in the Appendix. 

\section{Related Work}
\citet{AbSo03b} and \citet{Abdu05a} explored the school choice model with rigid racial and gender quotas, imposing limits on the number of admitted students from specific groups. Schools commonly set both maximum and minimum quotas for each student type \citep{HYY13a, Koji12a, KoSo13a}. \citet{EHYY14a} developed the \emph{controlled school choice} model, analyzing the effects of treating diversity quotas as hard or soft bounds, while \citet{EcYe15a} investigated choice functions that satisfy substitutability.

These studies assume that each student belongs to a single type. However, real-world programs in Brazil, Chile, Israel, and India consider overlapping types \citep{AyTu16a, BCC+19a, CEE+19a, KHIY17a, GNKR19a}. When individuals have multiple types, two main conventions determine seat allocation: the \textit{one-for-all} and \textit{one-for-one} conventions \citep{SoYe22a}.

In the one-for-all convention, individuals occupy reserved seats for all qualifying types \citep{Aziz19b, AGS20a, GNKR19a}, but optimally meeting diversity goals under this convention is NP-hard \citep{BFIM10a}. The one-for-one convention, on the other hand, assigns an individual to a single reserved seat, either based on strict preferences over types \citep{AyTu16a, KHIY17a}, fixed tie-breaking \citep{BCC+19a, CEE+19a}, or dynamically updated priorities \citep{EHYY14a}.

In these cases, the decision on which type a student should use is often made sequentially, which may not maximize diversity when students have overlapping types. Similar approaches have been employed by \citet{KoSo13a, KS16a}, \citet{AyBo20a}, and \citet{AyTu20a, AyTu20b}. \citet{SoYe22a} introduced the concept of “smart reserves” in controlled school choice, offering a more flexible approach to type selection. \citet{AzSu21a} expanded on this by proposing a new matching model called matching with multi-rank reserves, providing a comprehensive framework to address various diversity constraints.

Type combinations were previously studied by \citet{AGS20a} under the one-for-all convention, where they derived new quotas for type combinations from original type quotas. However, this method does not fully maximize diversity relative to the original quotas, as it focuses on the new quotas for type combinations, neglecting the original type quotas.

Our new algorithms, based on the ranked reservation graph, build on the concept of rank-maximal matching \citep{IKM+06a}. Recent works by \citet{Pete2022a} and \citet{HMSS21a} have applied rank-maximal matching to elicit agent preferences in the house allocation model, which deals with allocating indivisible objects among agents.

%%%%%%%%%%%%%%%%%%%%%%%%%%%%%%%
% section : model 
%%%%%%%%%%%%%%%%%%%%%%%%%%%%%%%
\section{Model}
In this paper, we focus on how a single school selects students based on its diversity goals while adhering to a priority order, as explored in recent works~\citep{AzSu21b, SoYe22a, HKYY22a, YHK+23a}. In the last section, we will demonstrate how this selection procedure can be seamlessly incorporated into the generalized deferred acceptance algorithm \citep{HaMi05a}, allowing for the accommodation of multiple schools.

Diversity goals often take the form of minimum and maximum quotas, where schools first aim to fulfill minimum quotas, followed by maximum quotas whenever possible. This approach is known as \emph{dynamic priority} \citep{EHYY14a}, where precedence is given to students who contribute to achieving diversity goals, even if the resulting set of selected students does not strictly follow the school’s priority order over students.

\begin{figure}[h]
    \begin{center}
\begin{tikzpicture}
[scale=6]
\draw[-] (0,0) -- (1.0,0);
\foreach \x / \xtext in {0 / 0, 0.5 / minimum, 1/ maximum}
    \draw[thick] (\x,0.5pt) -- (\x,-0.5pt) node[below] {\xtext};
\draw (0.25, 0.5pt) node[above] {Rank 1};
\draw (0.75, 0.5pt) node[above] {Rank 2};
\end{tikzpicture}
\end{center}
\vspace{-1em}
\caption{An interpretation of minimum and maximum quotas for type $t$ as two ranks of quotas is as follows. Rank 1 corresponds to the scenario where the number of matched students of type $t$ does not exceed the minimum quota. Rank 2 applies when the number of matched students surpasses the minimum quota but does not exceed the maximum quota.}
\label{fig:rank}
\end{figure}
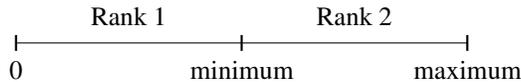

Minimum and maximum quotas can be translated into \emph{two ranks of quotas}, as illustrated in Figure~\ref{fig:rank}. These ranks represent the importance of each quota, with smaller ranks indicating higher importance. Schools prioritize filling quotas with smaller ranks first, consistent with the concept of dynamic priority.
This framework of multiple ranks provides a flexible way to model diverse constraints, including principles such as egalitarianism and proportionality. For further discussion on how such constraints are handled in school choice, we refer readers to \citep{AGSY20a, AzSu21b}.

An instance consists of a tuple \( I = (S, q, \succ, T, \eta) \), where \( S \) denotes a set of students, \( q \) represents the school capacity, \( \succ \) is a strict priority order over \( S \), and \( T \) denotes a set of types. With a slight abuse of notation, we use \( T(s) \) to denote the subset of types associated with student \( s \in S \). The parameter \( \eta \) represents a set of \emph{ranked quotas}, where \( \eta_{t}^j \) denotes the quota for type \( t \) at rank \( j \). For simplicity, we consider no more than two ranks of quotas in our examples throughout this paper. 

Let \( U \subseteq 2^T \) represent the set of type combinations, which are referred to as \emph{groups}. Each student \( s \) belongs to a unique group, denoted by \( U(s) \). We consider only the groups associated with the students in \( S \), so \( |U| \leq |S| \).

Next, we illustrate the issue of imbalance across groups using Example~\ref{example:issue}. This aspect, overlooked in previous studies, underscores the need for further consideration.

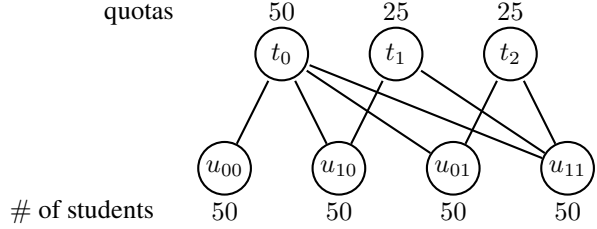
\begin{figure}[tb]
\begin{center}

\begin{tikzpicture}[-,>=stealth',shorten >=1pt,auto,node distance=2cm,
thick,main node/.style={circle,fill=white!20,draw,minimum size=0.7cm,inner sep=0pt]}, scale=0.38]

\node[main node] at (-6, 0) (1)  {$u_{00}$};
\node[main node] at (-2, 0) (2)  {$u_{10}$};
\node[main node] at (2, 0) (3)  {$u_{01}$};
\node[main node] at (6, 0) (4)  {$u_{11}$};

\node[] at (-11, -1.5) {$\#$ of students};
\node[] at (-6, -1.5) {$50$};
\node[] at (-2, -1.5) {$50$};
\node[] at (2, -1.5) {$50$};
\node[] at (6, -1.5) {$50$};

\node[main node] at (-4, 4) (t0)  {$t_0$};
\node[main node] at (0, 4) (t1)  {$t_1$};
\node[main node] at (4, 4) (t2)  {$t_2$};

\node[] at (-9, 5.4) {quotas};
\node[] at (-4, 5.5) {$50$};
\node[] at (0, 5.5) {$25$};
\node[] at (4, 5.5) {$25$};

% \node[text width=3cm] at (11,2) {rank $1$, type $t_1$};
%  \node[text width=3cm] at (11,0) {rank $1$, type $t_2$};
%    \node[text width=3cm] at (11,-2) {rank $2$, type $t_3$};

\draw [] (1) -- (t0);
\draw [] (2) -- (t1);
\draw [] (2) -- (t0);
\draw [] (3) -- (t2);
\draw [] (3) -- (t0);
\draw [] (4) -- (t1);
\draw [] (4) -- (t2);
\draw [] (4) -- (t0);

\end{tikzpicture}
\end{center}
\caption{Three types and four groups in Example~\ref{example:issue}.}
\label{fig:issue}
\end{figure}

\begin{example}
\label{example:issue}
Consider a school \( c \) with a capacity of 100 seats, which includes minimum quotas of 25 seats each for two types, \( t_1 \) and \( t_2 \). Additionally, the school reserves 50 seats for a general type \( t_0 \), available to any student. We define four groups as follows: \( u_{00} \), \( u_{10} \), \( u_{01} \), and \( u_{11} \), where the subscripts indicate the presence (1) or absence (0) of types \( t_1 \) and \( t_2 \). Specifically, \( u_{10} \) includes type \( t_1 \) but not \( t_2 \), \( u_{01} \) includes type \( t_2 \) but not \( t_1 \), and \( u_{11} \) includes both types, as depicted in Figure~\ref{fig:issue}. Assume there are 50 students in each group, and the priority ordering for students is \( u_{00} \succ u_{10} \succ u_{01} \succ u_{11} \). Here, we focus on how many students from each group will be chosen, rather than on which particular students will be selected.

The algorithms by \citet{SoYe22a} and \citet{AzSu21a} produce the same outcome: 50 students from \( u_{00} \), 25 students from \( u_{10} \), 25 students from \( u_{01} \), and 0 students from \( u_{11} \) are selected. 
While the minimum quotas for types \( t_1 \) and \( t_2 \) are met, the outcome is highly inequitable for students from \( u_{11} \), as none of them are chosen.
\end{example}

\section{Graph Structure}
In this section, we introduce two graph structures that underpin our results and algorithms. We begin by revisiting the \emph{ranked reservation graph} from previous work, followed by presenting a more compact representation using a flow network structure.

%%%%%%%%%%%%%%%%%%%%%%%%%%%%%%%%%%%%%%%%%%%%%
% section : Ranked Reservation Graph
%%%%%%%%%%%%%%%%%%%%%%%%%%%%%%%%%%%%%%%%%%%%%
\subsection{Ranked Reservation Graph}
\citet{IKM+06a} studied \emph{ranked bipartite graphs}, where edges are assigned ranks corresponding to students’ priorities. \citet{AzSu21a} introduced \emph{ranked reservation graphs} to represent diversity goals within the model of multiple ranks of quotas, where the ranks correspond to the quotas rather than to the priorities.
% A ranked reservation graph is a specialized type of ranked bipartite graph, in which all edges incident to the same reserved seat share the same rank.

Given an instance $I = (S, q, \succ, T, \eta)$, the corresponding ranked reservation graph $G = (S \cup V, E)$ is a bipartite graph with one set of vertices $S$ representing students and another set $V$ representing reserved seats.

Each reserved seat $v_{t,i}^j$ is associated with a rank $j$, a type $t$, and an index $i$. The indices distinguish between reserved seats of the same rank $j$ and type $t$. For rank $j$ and type $t$, there are $\eta_{t}^j$ reserved seats.
Additionally, a general type $t_0$ is introduced, available to all students. There are $q$ reserved seats for type $t_0$ with the largest rank.

An edge $(s, v_{t,i}^j)$ is added between student $s$ and reserved seat $v_{t,i}^j$ if the student has type $t$. Each edge $(s, v_{t,i}^j)$ is associated with the \emph{rank} $j$, corresponding to the rank of the reserved seat $v_{t,i}^j$. The set of all edges $E$ can be partitioned into $r$ categories: $E = E^1 \cup \cdots \cup E^r$, where $E^j$ denotes the set of edges with rank $j$, and $r$ represents the largest rank.

\begin{definition}[Matching]
\label{def:matching}
    Given a graph $G$, a matching $M$ is a set of edges in which each vertex appears in at most one edge of the matching. 
\end{definition}

\begin{definition}[Signature]
    \label{def:signature}
Given a ranked reservation graph $G$, the \emph{signature} of a matching $M$, denoted as $\rho(M)$, is a tuple of integers $\langle x_1, x_2, \ldots, x_r \rangle$, where each element $x_i$ represents the number of matched edges of rank $i$ in $M$.
\end{definition}

The signatures of two matchings are compared in a lexicographical manner. Specifically, a matching $M'$ with signature $\rho(M')$ $=$ $\langle x_1$, $\cdots$, $x_r \rangle$ is \emph{strictly better} than a matching $M''$ with signature $\rho(M'')$ $=$ $\langle y_1$, $\cdots$, $y_r \rangle$ if there exists an index $1 \leq k \leq r$ such that for all indices $i$ where $1 \leq i < k$, $x_i = y_i$ and $x_k > y_k$. A matching $M'$ is \emph{weakly better} than a matching $M''$ if $M''$ is not strictly better than $M'$.

\begin{definition}[Rank Maximality]
    \label{def:rank_maximality}
 Given a ranked reservation graph $G$, a matching $M$ of size at most $q$ is considered \emph{rank-maximal} if it is weakly better than any other matching of the same size.
\end{definition}
The size requirement in the definition of rank maximality is necessary because the total number of reserved seats may exceed the school's capacity when including $q$ reserved seats for the general type $t_0$.

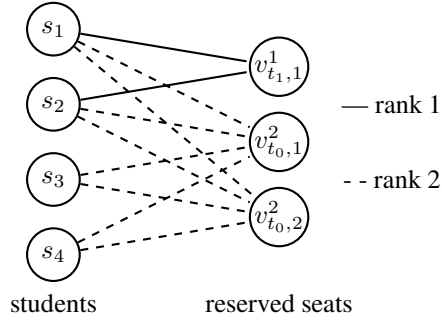
\begin{figure}[tb]
\begin{center}
\scalebox{1}{
\begin{tikzpicture}[-,>=stealth',shorten >=1pt,auto,node distance=3cm,
thick,main node/.style={circle,fill=white!20,draw,minimum size=0.7cm,inner sep=0pt]}, scale=0.5]
\node (students) at (0,-4.3) {students};
\node[main node] at (0, 3) (1)  {$s_1$};
\node[main node] at (0, 1) (2)  {$s_2$};
\node[main node] at (0, -1) (3)  {$s_3$};
\node[main node] at (0, -3) (4)  {$s_4$};
\node (seats) at (6,-4.3) {reserved seats};
\node[main node] at (6, 2) (d1)  {$v_{t_1, 1}^1$};
\node[main node] at (6, 0) (d2)  {$v_{t_0, 1}^2$};
\node[main node] at (6, -2) (d3)  {$v_{t_0, 2}^2$};
\node (A) at (9,1) {--- rank 1};
\node (B) at (9,-1) {- - rank 2};

\draw[] (1) -- (d1);            
\draw[dashed] (1) -- (d2);
\draw[dashed] (1) -- (d3);
\draw[] (2) -- (d1);
\draw[dashed] (2) -- (d2);
\draw[dashed] (2) -- (d3);
\draw[dashed] (3) -- (d2);
\draw[dashed] (3) -- (d3);
\draw[dashed] (4) -- (d2);
\draw[dashed] (4) -- (d3);
\end{tikzpicture}
}
\end{center}
\caption{A ranked reservation graph for Example~\ref{example:instance}.
Create one reserved seat $v_{t_1, 1}^1$ of rank $1$ for type $t_1$ and two reserved seats $v_{t_0, 1}^2$ and $v_{t_0, 2}^2$ of rank $2$ for type $t_0$. 
Solid lines incident to vertex $v_{t_1, 1}^1$ have rank $1$ and dashed lines incident to vertices $v_{t_0, 1}^2$ and $v_{t_0, 2}^2$ have rank $2$.}
\label{fig:example}
\end{figure}

\begin{example}
    \label{example:instance}
    Consider four students $S$ $=$ $\{s_1$, $s_2$, $s_3$, $s_4\}$ and two types: a privilege type $t_1$ and a general type $t_0$. Students $s_1$ and $s_2$ belong to type $t_1$, while all students are associated with type $t_0$. A minimum quota of 1 is imposed on type $t_1$, and the school has a capacity of 2, with the priority order $s_4$ $\succ$ $s_3$ $\succ$ $s_2$ $\succ$ $s_1$.

    The ranked reservation graph, depicted in Figure~\ref{fig:example}, uses solid lines to represent rank 1 edges and dashed lines for rank 2 edges. The matching $M_1$ $=$ $\{(s_1$, $v_{t_1, 1}^1)$, $(s_2$, $v_{t_0, 1}^2)\}$ has a signature $\rho(M_1)$ $=$ $\langle 1$, $1 \rangle$ and is rank-maximal. In contrast, the matching $M_2$ $=$ $\{(s_3$, $v_{t_0, 1}^2)$, $(s_4$, $v_{t_0, 2}^2)\}$ has a signature $\rho(M_2)$ $=$ $\langle 0$, $2 \rangle$.
\end{example}

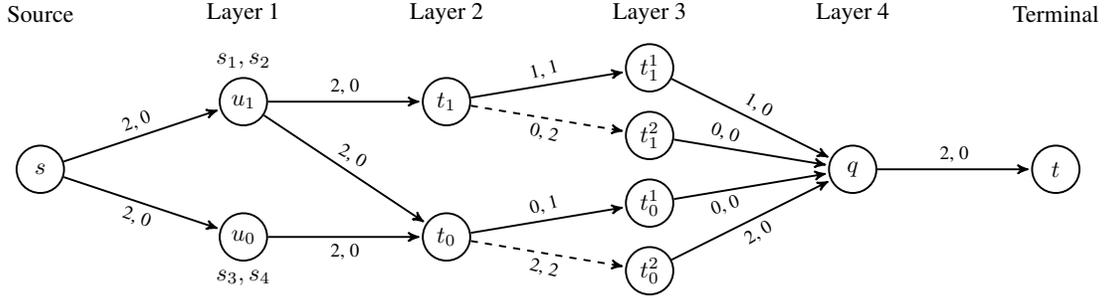
\begin{figure*}[htb]
\begin{center}
\scalebox{0.9}{ % Adjust this value to scale the whole diagram
\begin{tikzpicture}[
    -,>=stealth',
    shorten >=1pt,
    auto,
    node distance=2cm,
    el/.style = {inner sep=2pt, align=left, sloped},
    thick,
    main node/.style={circle,fill=white!20,draw,minimum size=0.7cm,inner sep=0pt}
]

% Nodes
\node (source) at (-3, 2.3) {Source};
\node [main node] (S) at (-3,0) {$s$};

\node (l1) at (0, 2.3) {Layer 1};
\node (s1) at (0, 1.6) {$s_1, s_2$};
\node (s2) at (0, -1.6) {$s_3, s_4$};
\node[main node] at (0, 1) (u1) {$u_1$};
\node[main node] at (0, -1) (u0) {$u_0$};

\node (l2) at (3, 2.3) {Layer 2};
\node[main node] at (3, 1) (t1) {$t_1$};
\node[main node] at (3, -1) (t2) {$t_0$};

\node (l3) at (6, 2.3) {Layer 3};
\node[main node] at (6, 1.5) (t11) {$t_1^1$};
\node[main node] at (6, 0.5) (t12) {$t_1^2$};
\node[main node] at (6, -0.5) (t21) {$t_0^1$};
\node[main node] at (6, -1.5) (t22) {$t_0^2$};

\node (l4) at (9, 2.3) {Layer 4};
\node [main node] (Q) at (9,0) {$q$};

\node (terminal) at (12, 2.3) {Terminal};
\node [main node] (T) at (12,0) {$t$};

% Edges
\draw[->] (S) -- (u1)  node[midway, above, rotate=20, scale=0.8] {2, 0}; 
\draw[->] (S) -- (u0)  node[midway, below, rotate=-20, scale=0.8] {2, 0}; 
\draw[->] (u1) -- (t1)  node[midway, above, scale=0.8] {2, 0}; 
\draw[->] (u1) -- (t2)  node[midway, above, rotate=-30, scale=0.8] {2, 0}; 
\draw[->] (u0) -- (t2)  node[midway, below, scale=0.8] {2, 0};  
\draw[->] (t1) -- (t11) node[midway, above, rotate=15, scale=0.8] {1, 1};
\draw[dashed, ->] (t1) -- (t12) node[midway, below, rotate=-15, scale=0.8] {0, 2};
\draw[->] (t2) -- (t21) node[midway, above, rotate=15, scale=0.8] {0, 1};
\draw[dashed, ->] (t2) -- (t22) node[midway, below, rotate=-15, scale=0.8] {2, 2};
\draw[->] (t11) -- (Q) node[midway, above, rotate=-30, scale=0.8] {1, 0};
\draw[->] (t12) -- (Q) node[midway, above, rotate=-20, scale=0.8, pos=0.3] {0, 0};
\draw[->] (t21) -- (Q) node[midway, below, rotate=20, scale=0.8, pos=0.3] {0, 0};
\draw[->] (t22) -- (Q) node[midway, below, rotate=30, scale=0.8] {2, 0};
\draw[->] (Q) -- (T) node[midway, above, scale=0.8] {2, 0};

\end{tikzpicture}
}
\end{center}
\caption{The flow network corresponding to Example~\ref{example:instance} consists of four layers, plus a source node and a terminal node. Each edge is labeled with $(c, w)$ where $c$ denotes the edge capacity and $w$ denotes the cost. Only edges between Layer 2 and Layer 3 have non-zero costs, representing their ranks. For example, dashed lines represent a cost of 2.}
\label{fig:network:example}
\end{figure*}

\subsection{Flow Network}
In this paper, we introduce a more compact graph structure that partitions agents into distinct \emph{groups} based on the types with which each agent is associated. Recall that each distinct type combination is referred to as a group $u\in U$.

Formally, given an instance $I = (S, q, \succ, T, \eta)$, the corresponding flow network $F = (V, E)$ is defined as a directed graph with capacities $c$ and costs (or weights) $w$. The flow network is organized into four layers, along with two additional vertices: a source vertex $s$ and a sink vertex $t$.

\begin{itemize}
 \item \emph{First Layer}: This layer consists of $|U|$ nodes, each corresponding to a group $u \in U$. A directed edge is created from the source vertex $s$ to each node $u$ in this layer, with a capacity equal to $|S_u|$ (the number of students from group $u$) and a weight of 0.
    \item \emph{Second Layer}: This layer contains $|T|$ nodes, each representing a type $t \in T$. For every group $u$ that includes type $t$, a directed edge is added from the node $u$ in the first layer to the node $t$ in the second layer. The capacity of this edge is 
    $|S_t \cap S_u|$ (the number of students with type $t$ from group $u$), and its weight is 0.
    \item \emph{Third Layer}: This layer consists of $|T| \times r$ nodes, where each node $t^i$ corresponds to a reserved quota $\eta_t^i$ with rank $i$ (with the largest rank being $r$). An edge is drawn from each node $t$ in the second layer to each node $t^i$ in the third layer, with a capacity of $\eta_t^i$ and a weight of $i$.
    \item \emph{Fourth Layer}: This layer includes a single node, denoted as $Q$, representing the school capacity. Each node $t^i$ in the third layer has a directed edge to the node $Q$, with a capacity of $\eta_t^i$ and a weight of 0.
    \item Finally, a directed edge is added from the node $Q$ to the sink vertex $t$, with a capacity equal to $q$ (the school capacity) and a weight of 0.
\end{itemize}

Note that only edges between Layer 2 and Layer 3 have non-zero costs, corresponding to the ranks of the quotas.

\begin{definition}[Flow]
\label{def:flow}
A \emph{flow} in a network $F$ is a function $f: V \times V \to \mathbb{R}$ that satisfies the following two properties:
\begin{enumerate}
    \item \textbf{Capacity Constraint}: For any $u, v \in V$, 
    \[
        0 \leq f(u, v) \leq c(u, v),
    \]
    where $c(u, v)$ denotes the capacity of the edge $(u, v)$.
    % from $u$ to $v$.
    \item \textbf{Flow Conservation}: For all $u \in V \setminus \{s, t\}$,
    \[
        \sum_{v \in V} f(v, u) = \sum_{v \in V} f(u, v).
    \]
    % where $s$ and $t$ are the source and sink vertices, respectively.
\end{enumerate}
\end{definition}

\begin{definition}[Minimum Cost Maximum Flow]
\label{def:maximum:flow}
The \emph{value} of a flow $f$ is defined as
\[
    |f| = \sum_{v \in V} f(s, v).
\]
% where $s$ is the source vertex.
The \emph{cost} of a flow $f$ is defined as
\[
    \text{Cost}(f) = \sum_{(v, u) \in E} w(v, u) \times f(v, u).
\]
% where $w(v, u)$ denotes the weight (or cost) associated with the edge from $v$ to $u$.
A \emph{minimum cost maximum flow} is a flow that maximizes the flow value while minimizing the total cost among all flows that achieve this maximum value.
\end{definition}

\begin{example}
\label{example:net_work}
Figure~\ref{fig:network:example} depicts the flow network of Example~\ref{example:instance}. Consider the matching $M_1$ $=$ $\{(s_1$, $v_{t_1, 1}^1)$, $(s_2$, $v_{t_0, 1}^2)\}$ with a signature $\rho(M_1) = \langle 1, 1 \rangle$ in the ranked reservation graph. It corresponds to a minimum cost maximum flow where two students $s_1$ and $s_2$ are sent from $s$ to $u_1$, student $s_1$ is then sent through $u_1$ $\rightarrow$ $t_1$ $\rightarrow$ $t_1^1$ $\rightarrow$ $q$ $\rightarrow$ $t$ and student $s_2$ is then sent through $u_1$ $\rightarrow$ $t_2$ $\rightarrow$ $t_0^2$ $\rightarrow$ $q$ $\rightarrow$ $t$.
\end{example}

\begin{restatable}{theorem}{TheoremGraph}
\label{theo:graph:equa}
Given an instance $I$, let $G$ be the corresponding ranked reservation graph, and $F$ be the associated flow network. A matching $M$ is rank-maximal in $G$ if and only if the corresponding flow $f$ is a minimum-cost maximum flow in $F$.
\end{restatable}

% We provide the reduction from a ranked reservation graph to a flow network in the Appendix.

\begin{proof}
We first prove the ``only if'' direction, i.e., if \( M \) is a rank-maximal matching in \( G \), then the corresponding flow \( f \) is a minimum-cost maximum flow in \( F \). 
To establish this, we introduce the following notations for a rank-maximal matching \( M \) in the ranked reservation graph \( G \):

\begin{itemize}
    \item \( M_u \): The set of students from group \( u \) who are matched in \( M \).
    \item \( M_{u,t} \): The set of students from group \( u \) who are matched to reserved seats for type \( t \) in \( M \).
    \item \( M_t^i \): The set of students who are matched to reserved seats of type \( t \) with rank \( i \) in \( M \).
\end{itemize}

We then define the corresponding flow \( f \) in the flow network \( F \) as follows:
\begin{itemize}
    \item For each edge between the source node \( s \) and a group node \( u \) in the first layer, set the flow \( f(s, u) = |M_u| \).
    \item For each edge between a group node \( u \) in the first layer and a type node \( t \) in the second layer, set the flow \( f(u, t) = |M_{u,t}| \).
    \item For each edge between a type node \( t \) in the second layer and a rank node \( t^i \) in the third layer, set the flow \( f(t, t^i) = |M_t^i| \).
    \item For each edge between a rank node \( t^i \) in the third layer and the capacity node \( Q \) in the fourth layer, set the flow \( f(t^i, Q) = |M_t^i| \).
    \item Finally, for the edge between the capacity node \( Q \) in the fourth layer and the sink node \( t \), set the flow \( f(Q, t) = |M| \), where \( |M| \leq q \).
\end{itemize}
This construction ensures that the flow \( f \) correctly represents the matching \( M \) in the flow network \( F \). Note that the total number of matched students in the matching \( M \) and the flow \( f \) is identical.

We now demonstrate that the flow \( f \) is a maximum flow. There are two cases to consider:
\begin{itemize}
    \item If the number of students is less than the quota, i.e., \( |S| < q \), then all students are selected.
    \item If the number of students is greater than or equal to the quota, i.e., \( |S| \geq q \), then exactly \( q \) students are selected due to the capacity constraint between the node \( Q \) in the fourth layer and the sink node \( t \).
\end{itemize}
In both cases, the flow \( f \) saturates the capacity constraints, ensuring that it is a maximum flow.

For the sake of contradiction, assume that the flow \( f \) is not a minimum-cost maximum flow. This would imply the existence of an augmenting cycle with negative cost in the corresponding residual graph \citep{AMO93a}. Recall that only the edges between the second and third layers are associated with costs, and these costs reflect the ranks from the ranked reservation graph. Therefore, we could adjust the flow by replacing some edges of \( f \) with edges of lower cost (i.e., lower rank) along the augmenting cycle. Reversing the transformation from the matching to the flow would yield a new matching with a strictly larger signature than the original matching \( M \). This contradicts the assumption that \( M \) is rank-maximal. Hence, our assumption was false, and \( f \) must be a minimum-cost maximum flow.

Next, we prove the ``if'' direction. Suppose that the flow \( f \) is a minimum-cost maximum flow in \( F \). We can construct a matching \( M \) in \( G \) as follows:
\begin{itemize}
    \item For each edge between the source node \( s \) and a node \( u \) in the first layer, select a set of students from group \( u \) of size \( f(s, u) \) and match them to group \( u \) in \( M \), i.e., \( |M_u| = f(s, u) \).
    \item For each edge between a node \( u \) in the first layer and a node \( t \) in the second layer, select a subset of students from \( M_u \) of size \( f(u, t) \) and match them to type \( t \) in \( M \), i.e., \( |M_{u, t}| = f(u, t) \).
    \item For each edge between a node \( t \) in the second layer and a node \( t^i \) in the third layer, select a subset of students from \( M_{u, t} \) of size \( f(t, t^i) \) and assign them to type \( t \) with rank \( i \) in \( M \), i.e., \( |M_t^i| = f(t, t^i) \).
\end{itemize}

The number of students matched in \( M \) is at most the quota \( q \), as the capacity of the edge between node \( Q \) and the terminal node \( t \) is constrained by \( q \).

For the sake of contradiction, suppose that the matching \( M \) is not rank-maximal. Then, there must exist another matching \( M' \) with more edges of smaller ranks. This would imply the existence of an augmenting cycle with a negative cost in the residual graph, leading to a contradiction.

Thus, the matching \( M \) is rank-maximal, completing the proof of Theorem~\ref{theo:graph:equa}.
\end{proof}

\section{Desirable Properties of a Choice Function}
In this section, we define some desirable properties of \emph{choice functions} for schools. Intuitively, a choice function selects a subset of students from a pool of applicants. For example, in the classical deferred acceptance algorithm \citep{GaSh62a}, a school's choice function simply selects students based on its priority order, up to its capacity.
In this paper, we aim to design a choice function that balances achieving diversity goals with respecting school priorities.

Formally, given an instance \( I = (S, q, \succ, T, \eta) \), a choice function \( Ch: I \rightarrow 2^S \) returns a subset \( S^* \subseteq S \) of students. We assume the choice function satisfies the feasibility requirement: \( |Ch(I)| \leq q \).

The first property is a modest efficiency criterion commonly used in two-sided matching models \citep{GIKY+16a, HHKS+17a, KHIY17a, AGS20a}. A choice function is \emph{non-wasteful} if a student is excluded only when the school's capacity \( q \) is fully utilized. Here, we assume all applicants are acceptable to the school, or we can exclude all unacceptable students.

\begin{definition}[Non-wastefulness]
\label{def:non-waste}
A choice function satisfies non-wastefulness if $|Ch(I)| = \min(|S|, q)$.
\end{definition}

Maximal diversity, as introduced by \citet{SoYe22a}, was initially applied to scenarios with a single rank of quotas (i.e., maximum quotas for types). The core idea is that the selected students should be able to form a maximum matching within the reserved seats graph. \citet{AzSu21a} extended this concept to models with multiple ranks of quotas by proposing the use of rank-maximal matchings instead of maximum matchings. In essence, a choice function satisfies maximal diversity if it consistently selects a set of students that can be included in a rank-maximal matching.

\begin{definition}[Maximal Diversity]
\label{def:maximal_diversity}
A choice function $Ch(I)$ satisfies maximal diversity if, for the corresponding ranked reservation graph $G$, there exists a rank-maximal matching $M$ of size at most $q$ such that $S_M \subseteq Ch(I)$, where $S_M$ denotes the set of students matched in $M$.
\end{definition}

To promote a more balanced distribution among different groups, we introduce the concepts of \emph{selection ratio} and \emph{balanced representation}. For a given matching $M$, the selection ratio for group $u$ is defined as the proportion of students selected from group $u$, i.e., $\frac{|M_u|}{|S_u|}$, where $|M_u|$ is the number of students chosen from group $u$, and $|S_u|$ is the total number of students in that group.

Intuitively, a choice function achieves balanced representation if it selects a set of students that maximizes the minimum selection ratio across all groups, among the matchings that also satisfy maximal diversity.

\begin{definition}[Balanced Representation]
\label{def:balance}
A choice function satisfies \emph{balanced representation} if there exists a matching $M \in \mathbb{M}$ such that the set of students selected by the choice function is matched in $M$ and 
\[
M \in \argmax_{M' \in \mathbb{M}} \ \min_{u \in U} 
\frac{|M_u'|}{|S_u|}
\]
where $\mathbb{M}$ denotes the set of all matchings that adhere to maximal diversity, $M_u'$ represents the subset of students from group $u$ who are matched in $M'$, and $S_u$ denotes the set of all students belonging to group $u$.
\end{definition}

\begin{example}
\label{exam:balance}
Consider Example~\ref{example:issue} again. For a balanced representation matching, the minimum selection ratio is $50\%$ for all groups, where the number of matched students from each group is exactly 25. 
\end{example}

Alternative methods for designing the selection ratio exist beyond the percentage of selected agents from each group, and our proposed algorithms are still applicable to these methods. We next present a more general way to define the selection ratio in Definition~\ref{def:general_selection_ratio}, attributed to \citep{STW21a} who focused on a pairwise exchange problem.

% \begin{definition}[Selection Ratio]
% \label{def:selection_ratio}
% Given a matching $M$ in $G$, and two targets $\overline{\delta}_u$ and $\underline{\delta}_u$ with $\overline{\delta}_u \geq \underline{\delta}_u$, the selection ratio of group $u \in U$ with respect to $\overline{\delta}_u$ and $\underline{\delta}_u$ is defined as 
% \[
% \alpha(|M_u|, \overline{\delta}_u, \underline{\delta}_u) =  \frac{|M_u| - \underline{\delta}_u}{\overline{\delta}_u - \underline{\delta}_u} \quad \text{for } \overline{\delta}_u > \underline{\delta}_u.
% \]
% When $\overline{\delta}_u = \underline{\delta}_u$, we assume $\alpha(|M_u|, \overline{\delta}_u, \underline{\delta}_u) = -\infty$ if $|M_u| < \overline{\delta}_u$ and $\alpha(|M_u|, \overline{\delta}_u, \underline{\delta}_u) = \infty$ if $|M_u| \geq \overline{\delta}_u$.
% \end{definition}

\begin{definition}[General Selection Ratio]
\label{def:general_selection_ratio}
Given a matching $M$ in $G$, and two targets $\overline{\delta}_u$ and $\underline{\delta}_u$ with $\overline{\delta}_u \geq \underline{\delta}_u$, the selection ratio of group $u \in U$ with respect to $\overline{\delta}_u$ and $\underline{\delta}_u$ is defined as 
\[
\alpha(|M_u|, \overline{\delta}_u, \underline{\delta}_u) =  \frac{|M_u| - \underline{\delta}_u}{\overline{\delta}_u - \underline{\delta}_u} \quad \text{for } \overline{\delta}_u > \underline{\delta}_u.
\]
When $\overline{\delta}_u = \underline{\delta}_u$, we assume $\alpha(|M_u|, \overline{\delta}_u, \underline{\delta}_u) = -\infty$ if $|M_u| < \overline{\delta}_u$, and $\alpha(|M_u|, \overline{\delta}_u, \underline{\delta}_u) = \infty$ if $|M_u| \geq \overline{\delta}_u$.
\end{definition}

Depending on the choice of $\overline{\delta}_u$ and $\underline{\delta}_u$, different ideas of selection ratio can be implemented, such as proportional to group sizes or egalitarianism.
For more details, we refer the readers to \citep{STW21a}. For simplicity and clarity, we focus on Definition~\ref{def:balance} in this paper.

% \subsection{Justified Envy-freeness}
In the classical school choice problem without diversity constraints \citep{AbSo03b}, student $s$ is said to have envy towards student $s'$ if student $s$ prefers the school $c$ assigned to $s'$ and has a higher priority at school $c$ than $s'$. A matching is considered fair if no student has envy towards another. In our setting, we can reinterpret this concept: an unmatched student has envy towards another if the student has a higher priority. However, in general, this fairness concept is incompatible with non-wastefulness and maximal diversity, even if each student is associated with one type \citep{EHYY14a}.

A natural, less stringent version of fairness restricts envy to occur only within the same group of students.

\begin{definition}[Respect of Priorities for Same Groups]
\label{def:fair_type_comb}
A choice function respects priorities for students from the same group if, for any unmatched student $s$ and for any student $s' \in Ch(I)$ with $U(s') = U(s)$, it holds that $s' \succ s$.
\end{definition}

We now introduce the concept of justified envy-freeness, which is stronger than respect for priorities within the same group. This concept asserts that a student with higher priority is unmatched only if replacing a student with lower priority would compromise the criteria of maximal diversity or balanced representation.

\begin{definition}[Justified Envy-freeness]
\label{def:JEF}
Given an instance $I$, a choice function $Ch(I)$ satisfies justified envy-freeness if, for any unselected student $s \notin Ch(I)$, there does not exist a student $s' \in Ch(I)$ such that, if $S' = Ch(I) \cup \{s\} \setminus \{s'\}$, the following conditions hold: (i) $s \succ s'$, and (ii) $S'$ satisfies both maximal diversity and balanced representation.
\end{definition}

We next illustrate the difference between these concepts.

\begin{example}
\label{example:comparision:fairness}
Consider Example~\ref{example:instance} again. Matching $M_1 = \{(s_1, v_{t_1, 1}^1), (s_2, v_{t_2, 1}^2)\}$ respects the priorities of students within the same group. However, it does not satisfy justified envy-freeness. This is because student $s_4$, who has a higher priority, could potentially replace either $s_1$ or $s_2$ without violating the conditions of maximal diversity or balanced representation. 
On the other hand, matching $M_2 = \{(s_2, v_{t_1, 1}^1), (s_4, v_{t_0, 1}^2)\}$ satisfies justified envy-freeness while preserving both maximal diversity and balanced representation.
\end{example}

Although these properties are defined for the choice function of schools, we sometimes use the term loosely by stating that ``a matching'' satisfies a particular property instead of the choice function. This is because these properties can be framed such that there exists a matching that meets certain conditions in the corresponding ranked reservation graph, with the choice function returning the set of students who are matched in such a matching.

\section{Design of Choice Function}
In this section, we first introduce a straightforward choice function that uniquely satisfies the four key properties: maximal diversity, non-wastefulness, justified envy-freeness, and balanced representation. We then provide a high-level overview of how this choice function is implemented. The specific methods for implementation are dependent on the two graph structures we utilize, and we defer the technical details to the subsequent sections.

\begin{algorithm}[htb]
\begin{algorithmic}[scale=1]
    \REQUIRE An instance $I = (S, q, \succ, T, \eta)$
    \ENSURE A set of students $S^* \subseteq S$
\end{algorithmic}
\begin{algorithmic}[1]
      \caption{Maximum and Balanced Choice Function}
      \label{Alo:choice_func}
      \STATE Selected students $S^* \leftarrow \emptyset$ 
      \FOR{each $s \in S$ in the descending order of $\succ$}
        \IF{there exists a matching $M$ of size at most $q$ that satisfies maximal diversity and balanced representation, and includes all students in $S^* \cup \{s\}$}
            \STATE $S^* \leftarrow S^* \cup \{s\}$
        \ENDIF 
      \ENDFOR
      \RETURN $S^*$
\end{algorithmic}
\end{algorithm}

The new choice function is described in Algorithm~\ref{Alo:choice_func}. 
The algorithm iterates over each student $s$ based on the school's priority ordering $\succ$ and checks whether $s$ can be matched in a matching of size at most $q$ that achieves maximal diversity and balanced representation along with a set of previously selected students denoted by $S^*$. If so, $s$ is added to the set of chosen students $S^*$. Otherwise, the algorithm moves on to the next student. After traversing each student, the algorithm returns $S^*$ as output. The choice function is similar in spirit to \citep{SoYe22a} and \citep{AzSu21a}.

\begin{example}
\label{example:choice_func}
We illustrate Algorithm~\ref{Alo:choice_func} with Example~\ref{example:instance}. Recall that school $c$ imposes a minimum quota of 1 for type $t_1$ and has a capacity of 2. Balanced representation requires that one student is selected from each group. 
Algorithm~\ref{Alo:choice_func} checks each student in the decreasing order of school priority: $s_4, s_3, s_2, s_1$. Student $s_4$ is first added to $S^*$, since there exists a matching $M = \{(s_1, v_{t_1, 1}^1), (s_4, v_{t_0, 1}^2)\}$ satisfying maximal diversity and balanced representation. However, adding student $s_3$ to $S^*$ would violate balanced representation and maximal diversity, so $s_3$ is not selected. Student $s_2$ is then added to $S^*$, since the matching $M = \{(s_2, v_{t_1, 1}^1), (s_4, v_{t_0, 1}^2)\}$ satisfies maximal diversity and balanced representation. Student $s_1$ cannot be selected due to the capacity restriction. Thus, Algorithm~\ref{Alo:choice_func} returns $S^* = \{s_2, s_4\}$ as the outcome.
\end{example}

We next demonstrate that the choice function we propose is the unique one satisfying all four desirable properties. 
% This implies that any other choice function meeting these criteria will yield the same set of students as Algorithm~\ref{Alo:choice_func}.

\begin{restatable}{theorem}{TheoremChoiceUnique}
\label{theo:choice:unique}
  The choice function in Algorithm~\ref{Alo:choice_func} is the unique one which satisfies non-wastefulness, maximal diversity, justified envy-freeness and balanced representation. 
\end{restatable}

\begin{proof}
% The proof of Theorem~\ref{theo:choice:unique} consists of two parts.
We begin by proving that the choice function defined in Algorithm~\ref{Alo:choice_func} satisfies the four fundamental properties. Following this, we demonstrate that any choice function adhering to these four properties must yield the same set of selected students as Algorithm~\ref{Alo:choice_func}.

%%%%%%%%%%%%%%%%%%%%%%%%%%%%%%%%%
% proof of maximal diversity and balanced representation
%%%%%%%%%%%%%%%%%%%%%%%%%%%%%%%%%
\textbf{Maximal Diversity \& Balanced Representation:} Given an instance $I$, Algorithm~\ref{Alo:choice_func} iteratively selects a set of students $S^*$ who can be matched in a matching of size at most the school capacity $q$, ensuring that both maximal diversity and balanced representation are satisfied (line 3). Once the algorithm terminates, it returns a set of students who can be matched in a matching with these two properties.

\textbf{Non-wastefulness:}
Case i) Suppose the number of students $|S|$ is not smaller than the school capacity $q$, i.e., $|S| \geq q$. Recall that we create $q$ general seats of type $t_0$ with the largest rank, which are available to any student. Hence, there are enough seats for $q$ students to be matched.
Case ii) Suppose $|S| < q$. In a ranked reservation graph, a rank-maximal matching is also a maximum size matching of size at most $q$ \citep{AzSu21a}. Assume that not all students from $S$ are matched. We could then increase the size of the rank-maximal matching by finding an augmenting path, without exceeding the school capacity. However, this would lead to a contradiction, as the matching produced by the choice function is supposed to be a maximum matching. Therefore, all students in $S$ must be matched.
In both cases, we have $|Ch(I)| = \min(|S|, q)$, which implies non-wastefulness.

%%%%%%%%%%%%%%%%%%%%%%%%%%%%%%%%%
% proof of no envy
%%%%%%%%%%%%%%%%%%%%%%%%%%%%%%%%%
\textbf{Justified Envy-freeness:}
For the sake of contradiction, suppose Algorithm~\ref{Alo:choice_func} does not satisfy justified envy-freeness. By definition, this would mean there exists an unselected student $s \in S \setminus Ch(I)$ and another student $s' \in Ch(I)$ such that the set $S' = Ch(I) \cup \{s\} \setminus \{s'\}$ satisfies the following conditions: (i) $s \succ s'$, and (ii) $S'$ satisfies both maximal diversity and balanced representation. By the time student $s$ is rejected, denote the set of chosen students as $S''$. When student $s'$ is chosen, all students in $S''$ are still chosen, as $s'$ is checked after student $s$. Therefore, there exists a matching of size at most $q$ that satisfies maximal diversity and balanced representation, and includes $s$ along with all previously chosen students in $S''$ who have higher priority than $s$. However, this contradicts the assumption that $s$ was not selected by Algorithm~\ref{Alo:choice_func}. Therefore, the algorithm must satisfy justified envy-freeness.

\textbf{Uniqueness:} 
We now proceed to prove that the choice function in Algorithm~\ref{Alo:choice_func} is the unique one satisfying the four properties. 
Consider any other choice function $\lambda$ that satisfies all four properties, and let $\widetilde{S}$ denote the set of students selected by $\lambda$. 
Let $S^* = \{s_1, s_2, \dots, s_o\}$ represent the set of students chosen by Algorithm~\ref{Alo:choice_func}, ranked in decreasing order of school priority $\succ$. 
We will now iteratively prove that each student from $S^*$ must be included in $\widetilde{S}$, thereby demonstrating that $\widetilde{S} = S^*$.

Let $\alpha$ denote the max-min selection ratio, and let $\delta^* = (\delta_u)_{u\in U}$ represent the crucial vector in Definition~\ref{def:Max-min-ratio} corresponding to the max-min selection ratio $\alpha$, where $\delta_u^* = \lfloor \alpha \cdot |S_u| \rfloor$ for all $u \in U$.

If $\sum_{u\in U}\delta_u^* = q$, then the number of students selected from each group is fixed, and we must choose the top $\delta^*_u$ students from each group; otherwise, we would violate justified envy-freeness. This ensures that $\widetilde{S} = S^*$. 

Now suppose $\sum_{u\in U}\delta_u^* < q$. This implies that there must exist at least one group $u$ where the number of selected students exceeds the target $\delta_u^*$. 

\emph{Base Case:} 
Initialize $S^*$ to be empty. Any student $s'$ with $s' \succ s_1$ cannot be chosen by $\lambda$; otherwise, this would lead to a contradiction, as $s_1$ is the first student who can be included in a matching of size at most $q$ that achieves both maximal diversity and balanced representation.

For the sake of contradiction, suppose $s_1$ is not selected by $\lambda$, i.e., $s_1 \notin \widetilde{S}$. Let $f$ and $f'$ denote the flows corresponding to Algorithm~\ref{Alo:choice_func} and the choice function $\lambda$, respectively. Both flows satisfy maximal diversity (i.e., a minimum-cost maximum flow) and balanced representation (i.e., the number of matched students from each group $u$ is no less than the target $\delta_u^*$). The only difference is that $s_1$ is selected in $f$ but not in $f'$.

Any student from the same group as $s_1$ cannot be selected by the choice function $\lambda$; otherwise, by definition, $s_1$ would have justified envy towards that student. Therefore, no student from group $u(s_1)$ is selected in $\widetilde{S}$. This implies that there must be another group $u'$ where the number of selected students increases in $f'$ compared to $f$.

Consider the second and third layers in both flow networks. If we examine the undirected edges, two cases can arise: either group $U(s_1)$ and group $u'$ are connected, or they are not. i) In the first case, where $U(s_1)$ and $u'$ are connected, there exists an alternative path between them. We can increase the flow at $U(s_1)$ by 1 and decrease the flow at $u'$ by 1, while keeping all other flows unchanged. This adjustment maintains maximal diversity because the in-flow at each node in the third layer remains the same.
ii) In the second case, where $U(s_1)$ and $u'$ are not connected, we can still increase the flow at $U(s_1)$ by 1 and decrease the flow at $u'$ by 1, again keeping all other flows unchanged. If doing so were to violate maximal diversity, it would imply that the flow network $f$ is not a minimum-cost maximum flow, which is a contradiction. Hence, maximal diversity is still maintained in this case as well.
In both scenarios, decreasing the flow at $u'$ by 1 does not violate balanced representation because the flow network $f$ already satisfies this property.

Note that any student chosen from $u'$ does not have higher priority than $s_1$. This would violate justified envy-freeness, contradicting our assumption. Therefore, we must include $s_1$ in $\widetilde{S}$.

\emph{Inductive Step:} Assume that the students $s_1, s_2, \dots, s_k$ have already been included in $S^*$. Now, consider student $s_{k+1}$. For any student $s'$ with a priority between $s_k$ and $s_{k+1}$, if $s'$ were selected by $\lambda$, it would contradict the assumption that $s_{k+1}$ is the next student who can be included in a matching of size at most $q$ that satisfies both maximal diversity and balanced representation, along with the set $S^*$. Similar to the case with student $s_1$, if $s_{k+1}$ were not chosen, this would violate justified envyfreeness.

By induction, each student $s_1, s_2, \dots, s_o$ chosen by Algorithm~\ref{Alo:choice_func} must be included in $\widetilde{S}$. Since both choice functions achieve maximal diversity, the number of matched students must be the same, which implies that $S^* = \widetilde{S}$, proving that the choice function in Algorithm~\ref{Alo:choice_func} is the unique one satisfying the four properties.

This completes the proof of Theorem~\ref{theo:choice:unique}.
\end{proof}

\subsection{High Level Description of Implementation}
%
% Algorithm~\ref{Alo:choice_func} does not provide detailed implementation instructions, particularly concerning how to .

Next, we introduce two concepts that are critical for the effective implementation of Algorithm~\ref{Alo:choice_func}. Intuitively, an instance is considered \emph{valid} with respect to a target vector for groups if it allows for a matching that maximizes diversity while ensuring that the number of matched students from each group meets the specified targets.

\begin{definition}[Validity]
\label{def:validity}
An instance $I = (S, q, \succ, \eta)$ is considered \emph{valid} with respect to a target vector $\delta = (\delta_u)_{u \in U}$ if there exists a matching $M$ such that:
\begin{enumerate}
\item $M$ achieves maximal diversity, and
\item For all $u \in U$, $|M \cap S_u| \geq \delta_u$, where $S_u$ denotes the set of students from group $u$.
\end{enumerate}
\end{definition}

If we have an algorithm $\Gamma$ that can check in polynomial time whether an instance is valid with respect to a given target vector for groups, we can use binary search, invoking $\Gamma$, to compute the max-min selection ratio needed to achieve balanced representation, as shown shortly.

\begin{definition}[Max-min Ratio \& Crucial Vector]
\label{def:Max-min-ratio}
Given an instance $I$, let $\alpha$ denote the max-min selection ratio that ensures balanced representation. Let $\delta^* = (\delta_u)_{u\in U}$ be the crucial vector corresponding to the max-min selection ratio $\alpha$, where $\delta_u = \lfloor \alpha \cdot |S_u| \rfloor$ for all $u \in U$.
\end{definition}

The following Theorem~\ref{theo:validity} asserts that if an instance $I$ is valid with respect to the crucial vector (corresponding to balanced representation), then a matching exists that satisfies all four properties. This theorem provides a sufficient condition to guarantee the existence of such a matching.

\begin{restatable}{theorem}{TheoremValidity}
  \label{theo:validity}
  An instance $I$ is valid with respect to its crucial vector $\delta^*$ if and only if it admits a matching that satisfies non-wastefulness, maximal diversity, balanced representation, and justified envy-freeness.
\end{restatable}

\begin{proof}
If an instance $I$ admits a matching that satisfies all four axioms, then it is trivially valid with respect to the crucial vector by the definition of balanced representation.

Suppose that an instance is valid with respect to the crucial vector. We know that it admits a matching satisfying three axioms: non-wastefulness, maximal diversity, and balanced representation. We can then select students one by one to achieve justified envy-freeness while maintaining maximal diversity and balanced representation, as described in Algorithm~\ref{Alo:choice_func}. This guarantees that the final matching respects all four axioms.

This completes the proof of Theorem~\ref{theo:validity}.
\end{proof}

 We now present a high-level overview of the implementation of the choice function in Algorithm~\ref{Alo:choice_func} by addressing the following two key questions:
\begin{itemize}
\item The \textbf{first question} involves designing an efficient algorithm to check whether an instance is valid, which allows us to compute the max-min selection ratio among all groups using binary search.
\item The \textbf{second question} focuses on designing an efficient algorithm to verify the existence of a matching with the desired properties, while ensuring that a specific set of students is included (line 3). We can then repeat this procedure and update the matching accordingly to ensure justified envy-freeness, while maintaining both maximal diversity and balanced representation.
\end{itemize}

\section{Algorithm Design based on Rank-maximal Matching}

In this section, we propose two polynomial-time algorithms based on rank-maximal matching to solve the two technical questions in implementing Algorithm~\ref{Alo:choice_func}.
% : i) how to check whether an instance is valid w.r.t. a given vector $\eta$; and ii) given a valid instance that satisfies balanced representation, how to check whether there exists a  matching that satisfies both maximal diversity and balanced representation while including a certain set of students.

Given a matching $M$, an \emph{alternating path} is a set of edges that begins with an unmatched vertex and whose edges belong alternately to the matching and not to the matching. An \emph{augmenting path} is an alternating path that starts and ends on unmatched vertices. Given an augmenting path $P$, let $M \oplus P$ denote a new matching in which edges from $P \setminus M$ are matched while edges from $M \cap P$ are not matched.

\subsection{Checking Validity}

Algorithm~\ref{Alg:validity} solves the first question of checking whether a given instance is valid w.r.t. some target vector $\delta$
and it returns a rank-maximal matching for a yes-instance.

To begin with, the algorithm constructs the ranked reservation graph $G$ for the given instance $I$. Then, it computes a rank-maximal matching $M$ with a size up to the capacity $q$. The set of students matched in this initial matching is denoted as $S_M$ and the set of chosen students $\widetilde{S}$ is initialized as empty (line 1-3).

Next, the algorithm selects a subset of students from $S_M$. For each group $u$, it checks whether the number of matched students from that group in $S_M$ meets the target $\delta_u$. If it does, the algorithm picks exactly $\delta_u$ students from this group. If the number of matched students is less than $\delta_u$, all the students from the group in $S_M$ are chosen. These selected students are added to the set $\widetilde{S}$ (line 4-6).  

The algorithm then proceeds to adjust the matching $M$ to meet the validity requirements. For each undersubscribed group $u$ (where the number of matched students is less than $\delta_u$), the algorithm looks for an unmatched student $s$ from $S_u \setminus S_M$. It performs the following checks and updates:
\begin{itemize}
    \item It first checks if there is an alternating path with respect to $M$ that starts from $s$ and ends at a student $s'$ in $S_M \setminus \widetilde{S}$ (a student matched in $M$ but not yet included in $\widetilde{S}$). If such a path is found, $s$ is added to $\widetilde{S}$, and the matching is updated by applying $M \oplus P$ (line 9-10).
    \item Otherwise, the algorithm looks for an augmenting path with respect to $M$ that starts from $s$ and ends at an unmatched seat $v$ of rank $k$. It also checks if there is a student $s'$ from $S_M \setminus \widetilde{S}$ who is matched to a seat $v'$ of the same rank. If both conditions are satisfied, $s$ is added to $\widetilde{S}$, and the matching is updated by applying $M \oplus P$ and excluding the edge $(s', v')$ (line 11-12).
\end{itemize}
The procedure continues until the number of students of each group $u$ in $\widetilde{S}$ reaches the minimum target $\delta_u$ for that group. If it is not possible to achieve the targets, the algorithm returns NO-instance. Once all targets are successfully met, the algorithm returns the updated matching $M$.

\begin{algorithm}[tb]
\begin{algorithmic}[scale=1]
\REQUIRE An instance $I=$ $(S$, $q$, $\succ$ , $T$, $\eta$) and a target vector $\delta = (\delta_u)_{u\in U}$
\ENSURE a rank-maximal matching $M$ of size at most $q$ s.t. $\forall u\in U, |M_u| \geq \delta_u$ or no-instance
\end{algorithmic}
\begin{algorithmic}[1]
\caption{Checking Whether an Instance is Valid}
\label{Alg:validity}
 \STATE Construct a  ranked reservation graph $G$ and 
compute a rank-maximal matching $M$ of size at most $q$ 
 % in $G$ by the algorithm \citep{AzSu21b} 
 \STATE Denote the set of students matched in $M$ as $S_M$ 
 \STATE $\widetilde{S} \leftarrow \emptyset$
 \FOR{each group $u\in U$}
 \STATE Choose $\min(\delta_u, |S_{M,u}|)$ students from $S_{M, u}$ (the set of students from group $u$ in $S_M$) and add them to $\widetilde{S}$
 \ENDFOR
 \FOR{each group $u \in U$}
\WHILE{$|\widetilde{S}_u| < \delta_u$ \AND $\exists s \in S_u \setminus S_M$}
\IF {there exists an \emph{alternating path} $P$ w.r.t. $M$ that starts from $s$ and ends at $s' \in S_M \setminus \widetilde{S}$} 
\STATE $\widetilde{S} \leftarrow \widetilde{S} \cup \{s\}$, $M \leftarrow M \oplus P$ 
\ELSIF{ {there exist i) an \emph{augmenting path} $P$ w.r.t. $M$ that starts from $s$ and ends at some unmatched seat $v$ of rank $k$ and ii) a student $s' \in S_M \setminus \widetilde{S}$ who is matched to some seat $v'$ of rank $k$ in $M$}}
\STATE $\widetilde{S} \leftarrow \widetilde{S} \cup \{s\}$, $M \leftarrow M \oplus P$, $M \leftarrow M \setminus \{(s', v')\}$ 
% \COMMENT{Remove student $s'$ from the matching $M$}} 
\ENDIF
\ENDWHILE
\IF{$|\widetilde{S}_{M,u}| < \delta_u$}
\RETURN NO-instance
\ENDIF
\ENDFOR
\RETURN matching $M$ 
\end{algorithmic}
\end{algorithm}

\begin{restatable}{theorem}{TheoremReserveValidity}
\label{theo:reserve:valid}  Given an instance $I$ and a target vector $\delta$, 
Algorithm~\ref{Alg:validity} checks whether the instance $I$ is valid w.r.t. $\delta$ in time $O(r |E| \sqrt{|V|} + |E| |V|)$, 
where $r$, $V$ and $E$ denote the number of ranks, the number of nodes and the number of edges in the ranked reservation graph.
\end{restatable} 

\begin{proof}
Given an instance $I$, we first construct the corresponding ranked reservation graph $G$ and compute a rank-maximal matching $M$ of size at most $q$ using the algorithm from \citet{AzSu21a}. If, for each group $u \in U$, the number of matched students from group $u$ is at least the target $\delta_u$, then we have found a feasible matching, and the algorithm terminates.

Suppose, instead, that there is a group $u \in U$ where $|M_u| < \delta_u$ and there exists a student $s$ from group $u$ who is not matched in $M$. If no such student from group $u$ is found, the algorithm returns ``no-instance,'' indicating that there are not enough students in group $u$ to meet the target $\delta_u$.

Let $\widetilde{S}$ denote the set of students that need to be matched to meet the targets $\delta$. For each group $u$, define $\widetilde{S}_u$ as a set of students such that $|\widetilde{S}_u| = \min(\delta_u, |S_{M,u}|)$, where $|S_{M,u}|$ is the number of students from group $u$ who are matched in $M$. Thus, $\widetilde{S}_u$ includes the smaller of the target $\delta_u$ and the number of students from group $u$ currently matched in $M$.

Consider a new matching $M'$ that differs minimally from $M$, achieves maximal diversity, and includes student $s$. Define the symmetric difference between $M$ and $M'$ as $(M \setminus M') \cup (M' \setminus M)$. Since $s$ is unmatched in $M$ but matched in $M'$, there must be an alternating path $P$ that starts from $s$ and alternates between edges in $M$ and $M'$. Given that the ranked reservation graph is bipartite, this path $P$ must end at either a student $s'$ or a reserved seat $v$.

\begin{itemize}
    \item \textbf{Case i:} Suppose there exists an alternating path $P$ with respect to $M$ that starts from $s$ and ends at some student $s'$, who is matched in $M$ but not selected by $\widetilde{S}$. In this case, we can update the matching $M$ by pairing $s$ with an appropriate reserved seat and removing the pairing of $s'$ with that seat. Specifically, let $M^* = M \oplus P$, where the edges of $P \setminus M$ are added to $M$ and the edges of $M \cap P$ are removed. As a result, all students in $\widetilde{S} \cup \{s\}$ are matched in $M'$. Since all edges incident to the same reserved seat have the same rank, the new matching $M'$ retains the same signature as $M$.

    \item \textbf{Case ii:} Suppose there exists an augmenting path $P$ with respect to $M$ that starts from $s$ and ends at an unmatched seat $v$ of rank $k$. Additionally, suppose there exists another student $s'$ who is matched in $M$ but not selected by $\widetilde{S}$ and is matched to some seat $v'$ of rank $k$. In this scenario, we can update the matching by pairing $s$ with $v$ and removing the pairing of $s'$ with $v'$. Specifically, let $M^* = M \oplus (P \setminus \{(s', v')\})$, where the edges of $P \setminus M$ are added to $M$, the edges of $M \cap P$ are removed, and the edge $(s', v')$ is removed. The new matching $M'$ retains the same signature as $M$ because the edge $(s, v)$ of rank $k$ replaces the edge $(s', v')$ of the same rank. Consequently, all students in $\widetilde{S} \cup \{s\}$ are matched in $M'$, which maintains the rank-maximal property of $M$.
\end{itemize}

In both cases, the number of matched students in group $u$ increases by 1 while maintaining maximal diversity. We continue this procedure until all groups meet their respective targets $\delta_u$. If, at any point, we encounter a situation where the matching cannot be updated to meet the target vector, it implies that the instance does not admit a matching satisfying the target vector.

The first step is to compute a rank-maximal matching, which takes time $O(r |E| \sqrt{|V|})$ \citep{AzSu21a}. Each time we compute an alternating or augmenting path, the length of the path is bounded by the number of edges $O(|E|)$ in the ranked reservation graph. The total number of such checks is proportional to the number of students, which is bounded by the number of nodes $|V|$. Therefore, the total running time of the algorithm is $O(r |E| \sqrt{|V|} + |E| |V|)$. This completes the proof of Theorem~\ref{theo:reserve:valid}.
\end{proof}

\begin{example}
\label{exam:test_valid}
We illustrate how Algorithm \ref{Alg:validity} works through Example~\ref{example:instance}.  Suppose school $c$ imposes a minimum target of $1$ for both groups $u_1$ and $u_0$. First, we compute a rank-maximal matching of size at most $2$, say $M$ $=$ $\{$$(s_1$, $v_{t_0, 1}^2)$, $(s_2$, $v_{t_1, 1}^1)$$\}$. Both students $s_1$ and $s_2$ are from group $u_1$, and we need to select one of them. Let's choose $s_2$ from $S_M$ $=$ $\{s_1$, $s_2\}$ (where $S_M$ is the set of students matched in $M$), so we set $\widetilde{S}$ $=$ $\{s_2\}$.
For the undersubscribed group $u_0$, we select one unmatched student, say $s_3$, and check whether $s_3$ can be added to $\widetilde{S}$. Since there exists an alternating path from $s_3$ to $s_1$ through the seat $v_{t_0, 1}^2$, we can add $s_3$ to $\widetilde{S}$. Consequently, we update the matching to $M'=$ $\{$$(s_3$, $v_{t_0, 1}^2)$, $(s_2$, $v_{t_1, 1}^1)$$\}$. Thus, Example~\ref{example:instance} is valid and Algorithm \ref{Alg:validity} returns matching $M'$ as outcome.
\end{example}

\begin{algorithm}[tb]
\begin{algorithmic}[scale=1]
\REQUIRE An instance $I = (S, q, \succ, T, \eta)$, a critical vector $\delta^*$, a rank-maximal matching $M$ of size at most $q$ that is valid w.r.t. $\delta^*$, and the top $\delta^*_u$ students are matched.
\ENSURE a set of students $S^*$
\end{algorithmic}
\begin{algorithmic}[1]
\caption{Implementation of the Choice Function in Algorithm~\ref{Alo:choice_func} Based on Rank-maximal Matching}
\label{algo:rank:implementation}
\STATE Selected students $S^* \leftarrow \emptyset$ 
\FOR{each $s \in S$ in the descending order of $\succ$}
\IF{$s \in S_M$}
\STATE $S^* \leftarrow S^* \cup \{s\}$
\ELSE
\FOR {$s' \in S_M \cap S_u \setminus S^*$ s.t. $|S_M \cap S_u| > \delta_u^*$}
\IF {there exists an \emph{alternating path} $P$ w.r.t. $M$ that starts from $s$ and ends at $s'$} 
\STATE $S^* \leftarrow S^* \cup \{s\}$, $M \leftarrow M \oplus P$ 
\STATE \textbf{break}
\ELSIF{ {i) there exists an \emph{augmenting path} $P$ w.r.t. $M$ that starts from $s$ and ends at some unmatched seat $v$ of rank $k$ and ii) student $s'$ is matched to some seat $v'$ of rank $k$ in $M$}}
\STATE $S^* \leftarrow S^* \cup \{s\}$
\STATE $M \leftarrow M \oplus P$, $M \leftarrow M \setminus \{(s', v')\}$ 
\STATE \textbf{break}
% \COMMENT{Remove student $s'$ from the matching $M$}
\ENDIF
\ENDFOR
\ENDIF
\ENDFOR
\RETURN a set of students $S^*$

\end{algorithmic}
\end{algorithm}

\subsection{Implementation Based on Rank-maximal Matching}

We next provide a detailed implementation of the choice function in Algorithm~\ref{Alo:choice_func}, based on a rank reservation graph, as described in Algorithm~\ref{algo:rank:implementation}. 
The input consists of an instance $I$, a critical vector $\delta^*$, and a rank-maximal matching $M$ of size at most $q$ that is valid w.r.t. $\delta^*$ and the top $\delta^*_u$ students are matched for each $u\in U$. 

The algorithm begins by initializing the set $S^*$ as empty. The algorithm then examines each student $s$ in descending order of $\succ$ to determine whether $s$ can be included in $S^*$. If $s$ is already matched in $M$, $s$ is directly added to $S^*$ (line 3-4). 
If $s$ is not matched, the algorithm examines each student $s'$ who is matched in $M$, belongs to a group $u$ that has already met the target $\delta^*_u$, and is not yet included in $S^*$ as follows:
\begin{itemize}
    \item If there exists an alternating path $P$ with respect to $M$ that starts from $s$ and ends at $s'$, then $s$ is included in $S^*$, and update the matching $M$ to be $M \oplus P$ (line 7-9).
    \item If there exists an augmenting path $P$ with respect to $M$ that starts from $s$ and ends at an unmatched seat $v$ of rank $k$, and if student $s'$ is matched to another seat $v'$ of the same rank $k$ in $M$, $s$ is included in $S^*$, then $s$ is included in $S^*$, and update the matching $M$ to be $M \oplus P$ and exclude the edge $(s', v')$ (line 10-13).
\end{itemize}

\begin{restatable}{theorem}{TheoremRankImplementation}
    \label{theo:rank:implementation}
    Given a critical vector $\delta^*$, 
Algorithm~\ref{algo:rank:implementation} returns a matching satisfying maximal diversity, non-wastefulness, balanced representation and justified envy-freeness in time $O(|E| |V|)$ where $V$ and $E$ denote the number of nodes and the number of edges in the ranked reservation graph.
\end{restatable}

\begin{proof}
We start with a rank-maximal matching $M$ of size at most $q$ that is valid with respect to $\delta^*$, where the top $\delta^*_u$ students are matched. For each student $s$, processed in descending order of $\succ$, we check whether $s$ can be included in $S^*$ as follows.

If a student $s$ is ranked within the top $\delta^*_u$ positions according to $\succ$, then $s$ must be matched. For the sake of contradiction, suppose there is such a student $s$ who is not matched. If $s$'s seat is assigned to another student $s'$ from the same group, then $s$ has justified envy toward $s'$. If $s$'s seat is unassigned, then the matching does not satisfy balanced representation. Therefore, we must include $s$ in $S^*$.

If a student $s$ is not ranked within the top $\delta^*_u$ positions according to $\succ$, and suppose there exists a matching $M'$ that differs minimally from $M$, achieves maximal diversity, and includes student $s$ as well as $S^*$. Define the symmetric difference between $M$ and $M'$ as $(M \setminus M') \cup (M' \setminus M)$. Since $s$ is unmatched in $M$ but matched in $M'$, there must be an alternating path $P$ that starts from $s$ and alternates between edges in $M$ and $M'$. Given that the ranked reservation graph is bipartite, this path $P$ must terminate at either a student $s'$ or a reserved seat $v$.
\begin{itemize}
    \item \textbf{Case i:} Suppose there exists an alternating path $P$ with respect to $M$ that starts from $s$ and ends at some student $s'$, who is matched in $M$ but not selected by $S^*$, and whose group has met the target $\delta^*_u$. In this case, we can update the matching $M$ by pairing $s$ with an appropriate reserved seat and removing the pairing of $s'$ with that seat. Specifically, let $M^* = M \oplus P$, where the edges of $P \setminus M$ are added to $M$ and the edges of $M \cap P$ are removed. As a result, all students in $S^* \cup \{s\}$ are matched in $M'$. Since all edges incident to the same reserved seat have the same rank, the new matching $M'$ retains the same signature as $M$. Since $s'$ is from a group that has met the target $\delta^*_u$, excluding $s'$ does not violate the balanced representation.

    \item \textbf{Case ii:} Suppose there exists an augmenting path $P$ with respect to $M$ that starts from $s$ and ends at an unmatched seat $v$ of rank $k$. Additionally, suppose there exists another student $s'$ who is matched in $M$ but not selected by $S^*$, and is matched to some seat $v'$ of rank $k$, and whose group has met the target $\delta^*_u$. In this scenario, we can update the matching by pairing $s$ with $v$ and removing the pairing of $s'$ with $v'$. Specifically, let $M^* = M \oplus (P \setminus \{(s', v')\})$, where the edges of $P \setminus M$ are added to $M$, the edges of $M \cap P$ are removed, and the edge $(s', v')$ is removed. The new matching $M'$ retains the same signature as $M$ because the edge $(s, v)$ of rank $k$ replaces the edge $(s', v')$ of the same rank. Consequently, all students in $S^* \cup \{s\}$ are matched in $M'$, which maintains the rank-maximal property of $M$. Since $s'$ is from a group that has met the target $\delta^*_u$, excluding $s'$ does not violate the balanced representation.
\end{itemize}

This procedure aligns with the selection rule defined in the choice function used in Algorithm~\ref{Alo:choice_func}. For each student, we check whether there exists an alternating or augmenting path in time $O(|E|)$, and the total running time of the for loop is $O(|S||E|)$, which is bounded by $O(|V||E|)$.
\end{proof}

\begin{example}
\label{exam:rank:implementation}
We illustrate how Algorithm~\ref{algo:rank:implementation} works using Example~\ref{example:instance}. Suppose we start with the rank-maximal matching $M = \{(s_3, v_{t_0, 1}^2), (s_2, v_{t_1, 1}^1)\}$, which was produced by Algorithm~\ref{Alg:validity} as described in Example~\ref{exam:test_valid}. We first update $M$ to ensure that the top $\delta^*_u$ students are matched, resulting in $M = \{(s_2, v_{t_1, 1}^1), (s_4, v_{t_0, 1}^2)\}$.
We then check all students in the order $s_4, s_3, s_2, s_1$. Student $s_4$ is already matched and is added to $S^*$. Student $s_3$ cannot be added, as none of the required conditions are met. Student $s_2$ is already matched in $M$ and is added to $S^*$. Student $s_1$ is rejected, and the algorithm terminates.
\end{example}

\begin{algorithm}[htb]
\begin{algorithmic}[scale=1]
\REQUIRE An instance $I = (S, q, \succ, T, \eta)$
\ENSURE A crucial vector $\delta^* = (\delta_u^*)_{u \in U}$
\end{algorithmic}
\begin{algorithmic}[1]
    \caption{Computing a Crucial Vector based on Flow Network}
    \label{Algo:net_work:crucial_vector}
    \STATE Initialize $\delta_u^* \leftarrow 0$ for each group $u \in U$
    \STATE Initialize $\alpha_0 \leftarrow 0$ and $\alpha_1 \leftarrow 1$
    \WHILE{some $\delta_u^*$ differs from the last round}
        \STATE $\alpha \leftarrow (\alpha_0 + \alpha_1) / 2$
        \STATE $\delta_u^* \leftarrow \lfloor \alpha \cdot |S_u| \rfloor, \forall u \in U$
        \STATE Construct a flow network $F$. For each edge from $s$ to $u$ in the first layer, add a lower bound $\delta_u^*$
        \IF {there exists a minimum cost maximum flow}
            \STATE $\alpha_0 \leftarrow \alpha$ \COMMENT{Search between $\alpha$ and $\alpha_1$}
        \ELSE
            \STATE $\alpha_1 \leftarrow \alpha$ \COMMENT{Search between $\alpha_0$ and $\alpha$}
        \ENDIF
    \ENDWHILE
    \RETURN $\delta^*$
\end{algorithmic}
\end{algorithm}

The main challenge in designing algorithms based on the ranked reservation graph lies in addressing the first research question: computing a crucial vector. This can be achieved by employing Algorithm~\ref{Alg:validity} with binary search, which requires $\log(|S|)$ iterations. Each iteration runs in $O(r |E| \sqrt{|V|} + |E| |V|)$ time. Therefore, the total running time is $O(\log(|S|) r |E| \sqrt{|V|} + \log(|S|) |E| |V|)$.

\section{Algorithm Design Based on Flow Network}
In this section, we propose new algorithms to implement the choice function in Algorithm~\ref{Alo:choice_func} using flow networks. 
Compared to the methods based on a ranked reservation graph, 
these algorithms are more straightforward and much faster for large instances where the number of students significantly exceeds the number of groups.

For instance, to address the first question of checking validity, we can reformulate it as a minimum cost maximum flow problem by incorporating lower capacity constraints on certain edges. 
Specifically, given a target vector $\delta$ for groups, 
checking validity can be achieved by setting a lower bound of $\delta_u$ for each edge from the source node $s$ to each node $u$ corresponding to each group $u$ in the first layer in the flow network. Subsequently, we can apply any existing algorithm to solve the minimum cost maximum flow problem.

To distinguish it from the ranked reservation graph $G$, we denote the number of nodes and edges in the flow network $F$ as $n$ and $m$, respectively. Specifically, $n = O(|T| r + |U|)$ and $m = O(|U| |T| + r |T|^2)$, where $|U| \leq 2^{|T|} \leq |S|$.

\begin{restatable}{theorem}{TheoremNetworkValidity}
    \label{theo:network:validity}
Given an instance $I$ and a target vector $\delta$, let $F$ denote the corresponding flow network. Checking validity with respect to $\delta$ can then be done in time $O(m \log(n) (m + n \log(n)))$, where $m$ and $n$ denote the number of edges and nodes in the flow network $F$.
\end{restatable}

\begin{proof}
We have demonstrated that finding a matching that satisfies maximal diversity can be reduced to finding a minimum-cost maximum flow in Theorem~\ref{theo:graph:equa}. Checking validity with respect to the target vector $\delta$ is equivalent to verifying whether there exists a minimum-cost maximum flow with the additional condition that the number of students from each group meets or exceeds the specified minimum targets. This can be achieved by imposing a lower bound $\delta_u$ on the edges between the source node and each node $u$ in the first layer. If such a flow exists, the instance is valid with respect to the target vector. Note that the flow network problem with lower bounds can be converted into an equivalent flow network problem without any lower bounds, which can then be solved in strongly polynomial time $O(m \log(n) (m + n \log(n)))$ \citep{AMO93a}.
\end{proof}

\subsection{Computing a Crucial Vector}
We next proceed to compute a crucial vector in Algorithm~\ref{Algo:net_work:crucial_vector}.
Initialize the lower and upper selection ratios, $\alpha_0$ and $\alpha_1$, respectively. Set $\alpha$ to be the midpoint $(\alpha_0 + \alpha_1) / 2$. Define the target vector $\delta_u^*$ for each group $u$ as $\lfloor \alpha \cdot |S_u| \rfloor$.
If a minimum cost maximum flow can be found with this target vector, update the lower selection ratio to $\alpha$ and continue the search within the range $[\alpha, \alpha_1]$. If no feasible flow is found, update the upper selection ratio to $\alpha$ and search within the range $[\alpha_0, \alpha]$.
The algorithm terminates when the target values $\delta_u^*$ for all groups $u \in U$ stabilize and do not change between iterations.

\begin{restatable}{theorem}{TheoremNetworkCrucial}
\label{theo:network:crucial}  Given an instance $I$, 
Algorithm~\ref{Algo:net_work:crucial_vector} computes a crucial vector in time $O(m \log(n) (m + n\log(n))) \log(|S|)$ where $m$ and $n$ denote the number of edges and nodes in the flow network.
\end{restatable}

\begin{proof}
By Theorem~\ref{theo:network:validity}, we can determine whether an instance is valid with respect to a target vector by converting it into a minimum-cost maximum flow problem. Using this approach, we can apply binary search to compute the crucial vector, as demonstrated in Algorithm~\ref{Algo:net_work:crucial_vector}. This procedure requires at most $O(\sqrt{|S|})$ iterations, with each iteration running in strongly polynomial time $O(m \log(n) (m + n \log(n)))$.
\end{proof}

\subsection{Implementation Based on Flow Network}

We next proceed to the implementation of the choice function in Algorithm~\ref{Alo:choice_func} using a flow network approach. The algorithm first selects the top $\delta^*_u$ students from each group $S_u$ based on the priority order $\succ$. For each unselected student $s \in S \setminus S^*$, we then check whether $s$ can be included while still satisfying maximal diversity and balanced representation. This is done by verifying if there exists a minimum cost maximum flow with a lower bound of $|S_u'|$ on each edge from the source node to the node $u$, where $S' = S^* \cup \{s\}$. If such a flow exists, $s$ can be included in the selection.

\begin{algorithm}[ht]
\begin{algorithmic}[scale=1]
\REQUIRE An instance $I = (S, q, \succ, T, \eta)$ and a crucial vector $\delta^* = (\delta_u)_{u \in U}$
\ENSURE A set of students $S^*$
\end{algorithmic}
\begin{algorithmic}[1]
    \caption{Implementation of the Choice Function in Algorithm~\ref{Alo:choice_func} Based on Flow Network}
    \label{Alo:net_work:implementation}
    \STATE Initialize the selected students, $S^*$, as the union of the top $\delta_u$ students from each group $S_u$ according to $\succ$ for all $u \in U$.
    \STATE Construct a flow network $F$
    \FOR{each $s \in S \setminus S^*$ in descending order of $\succ$}
        \STATE Set $S' \leftarrow S^* \cup \{s\}$
        \STATE For each edge from $s$ to $u$ in the first layer, add a lower bound $|S_u'|$
        \IF {a minimum cost maximum flow exists}
            \STATE Update $S^* \leftarrow S^* \cup \{s\}$
        \ENDIF
    \ENDFOR
    \RETURN $S^*$
\end{algorithmic}
\end{algorithm}

\begin{restatable}{theorem}{TheoremNetworkImplementation}
    \label{theo:network:implementation}
    Given a crucial vector $\delta^*$, Algorithm~\ref{Alo:net_work:implementation}  returns a matching satisfying maximal diversity, non-wastefulness, balanced representation and justified envy-freeness in time $O(|S| m \log(n) (m + n\log(n)))$ where $m$ and $n$ denote the number of edges and nodes in the flow network.
\end{restatable}

\begin{proof}
\textbf{Maximal Diversity \& Non-wasfulness}
By the definition of crucial vector, we know there exists a minimum cost maximum flow which includes all students $S^*$. Due to the equivalence between the rank-maximal matching and minimum cost maximum flow in Theorem~\ref{theo:graph:equa}, we know it implies maximal diversity and non-wasfulness,

\textbf{Balanced Representation} During the process of Algorithm~\ref{Alo:net_work:implementation}, the number of matched students from each group does not decrease and thus it is weakly larger than $\delta^*_u$, which ensures  balanced representation.

\textbf{Justified Envy-freeness}
For the sake of contradiction, suppose an unmatched student $s\in S^*$ has justified envy towards $s' \in Ch(I)$. 
Algorithm~\ref{Alo:net_work:implementation} begins by selecting the top $\delta^*_u$ students $\widetilde{S}_u$ from $S_u$ and let 
$S^* = \bigcup_{u \in U} \widetilde{S}_u$. 
Thus $s \notin  \bigcup_{u \in U} \widetilde{S}_u$.
If $s\in S^*$ has justified envy towards $s' \in Ch(I)$, then there exists a minimum cost maximum flow in which $s$ is matched while $s'$ is not while the number of matched studetns from each group $u$ is weakly larger than $\delta^*_u$. However, Algorithm~\ref{Alo:net_work:implementation} would have selected student $s$ prior to $s'$, a contradiction.

\textbf{Running Time} Each iteration takes time $m \log(n) (m + n\log(n))$ and the number of iterations is bounded by  $|S|$. This completes the proof of Theorem~\ref{theo:network:implementation}.
\end{proof}

If we assume that the number of types and type combinations is bounded by a small constant, as is often the case in practical markets, then solving the problem based on the flow network structure is much faster than solving it using the ranked reservation graph.

\section{Multiple Schools}
In this section, we extend our choice function to address the problem of selecting students across a set of schools. An instance of this problem is formally represented as $I = (S, C, q, \succ_S, \succ_C, T, \eta)$, where $q = (q_c)_{c \in C}$ represents the capacity vector for each school, $\succ_S$ denotes the preference profile of students, $\succ_C$ denotes the priority profile of schools, and $\eta = (\eta_c)_{c \in C}$ represents the ranked quotas for each school.

For each school $c \in C$, we define the \emph{induced instance} for a subset of students $X \subseteq S$ applying to that school as $I_c(X) = (X, q_c, \succ_X, \succ_c, T, \eta_c)$, where $X$ is the subset of students applying to school $c$, $q_c$ is the capacity of school $c$, $\succ_X$ is the preference profile of students in $X$, $\succ_c$ is the priority profile of school $c$, $T$ represents the set of student types, and $\eta_c$ is the ranked quotas for school $c$. 

For brevity, we will denote the choice function for school $c$ simply as $Ch_c(X)$, instead of the more explicit form of $Ch_c(I_c(X))$.

\begin{algorithm}[h]
\begin{algorithmic}
\REQUIRE An instance $I = (S, C, q, \succ_S, \succ_C, T, \eta)$
\ENSURE A matching $M$
\end{algorithmic}
\begin{algorithmic}[1]
\caption{Generalized Deferred Acceptance (GDA)}
\label{alg:GDA}
\WHILE{some student makes a new proposal}
\STATE (Proposal) Each unmatched student applies to their most preferred school that has not yet rejected them.
\STATE (Selection) Let $S_c$ denote the set of students applying to school $c$ in this round and the students already matched to $c$ from the previous round. Each school $c$ selects a set of students $Ch_c(S_c)$ and rejects the remaining students. Update $M$ to reflect the current matching between students and schools.
\ENDWHILE
\RETURN Matching $M$
\end{algorithmic}
\end{algorithm}

The Generalized Deferred Acceptance (GDA) algorithm in Algorithm~\ref{alg:GDA}
 extends the classic deferred acceptance algorithm~\citep{HaMi05a}. The GDA operates as follows: unmatched students propose to their most preferred school, and schools select a subset of applicants based on a specified choice function. Students who are not chosen by a school are rejected and may propose to other schools in subsequent rounds. The process repeats until no further proposals are made. 
 
The choice function presented in Algorithm~\ref{Alo:choice_func} can be incorporated into the Generalized Deferred Acceptance (GDA) algorithm. However, its inclusion of balanced representation prevents it from satisfying the following property.

  \begin{definition}[Substitutability~\citep{KeCr82a}]
  \label{def:SUB}
  A choice function $Ch_c(X)$ satisfies substitutability if for all students $s_1, s_2 \in X$ and $Y \subseteq X$, if $s_2 \notin Ch_c(Y \cup \{s_2\})$, then 
  $s_2 \notin Ch_c(Y \cup \{s_1\} \cup \{s_2\})$.
  \end{definition}

Definition~\ref{def:SUB} essentially states that if a student $s_2$ is not selected from a set of applicants $Y \cup \{s_2\}$, then the presence of another student $s_1$ in the applicant pool should not affect this decision; $s_2$ will still not be selected when added to the set with $s_1$.

\begin{theorem}
    \label{theo:not:substuititube}
    The choice function in 
    Algorithm~\ref{Alo:choice_func} does not satisfy Substitutability.
\end{theorem}

\begin{proof}
Consider a school $c$ with a capacity of 4 seats. The student population consists of two types: $t_1$ and $t_2$, such that each student belongs to one of these types. Suppose the school has a single rank of quotas, i.e., the maximum quota for both types is 4.

Assume there are 5 students of type $t_1$, namely $s_{11}, s_{12}, s_{13}, s_{14}, s_{15}$, and 3 students of type $t_2$, namely $s_{21}, s_{22}, s_{23}$. The priority order of the school prefers all students of type $t_1$ over students of type $t_2$, as follows:
\[
s_{11}, s_{12}, s_{13}, s_{14}, s_{15}, s_{21}, s_{22}, s_{23}.
\]

Initially, the minimum selection ratio is given by $ \frac{2}{5} = 0.4 $, indicating that two students from each type are selected. Specifically, the selected students are $ s_{11}, s_{12}, s_{21}, s_{22} $, as returned by Algorithm~\ref{Alo:choice_func}.

Now, consider the addition of an extra student, $ s_{16} $, from type $ t_1 $ to the pool of applicants. In this scenario, the minimum selection ratio changes to $ \frac{1}{3} \approx 0.33 $, where three students from type $ t_1 $ and one student from type $ t_2 $ are now selected. The new selection includes $ s_{11}, s_{12}, s_{13}, s_{21} $, as determined by Algorithm~\ref{Alo:choice_func}.

This illustrates that the inclusion of $ s_{16} $ alters the composition of the selected students. Notably, the presence of $ s_{16} $ allows for the selection of $ s_{13} $, who was previously unselected. Therefore, the choice function fails to satisfy substitutability, as the selection of a student is contingent upon the presence of another student.

This completes the proof of Theorem~\ref{theo:not:substuititube}.
\end{proof}

Substitutability is a necessary condition for the Generalized Deferred Acceptance (GDA) algorithm to achieve the stability concept studied by \citep{HaMi05a}. Theorem~\ref{theo:not:substuititube} implies that GDA with our choice function cannot satisfy their stability concept.

% HM-Stability in Definition~\ref{def:HMstability} requires that matching $M$ needs to satisfy two conditions. First, for each agent $i$ (being either a student or a school), the assignment to agent $i$ equals the outcome of agent $i$'s choice function given $M(i)$ as input. In other words, each student / school is matched to some acceptable school / a set of acceptable students. 
% Second, there does not exist a pair of student and school who prefer to deviate from matching $M$ and be matched with each other. 

% \begin{definition}[HM-Stability]
% \label{def:HMstability}
% A matching $M$ satisfies HM-stability if  the following two conditions hold: 
% Let $M_c$ denote a set of students who are matched to $c$, and $M_s$ denote the school assigned to $s$ in the matching $M$.
% \begin{enumerate}
% \item (Individual rationality) for each student $s\in S$, $M_s$ is acceptable and for each school $c\in C$, $M_c = Ch_c(I_c(M_c))$.
% \item (No blocking pair)
% there exists no pair of student $s$ and school $c$ such that $c \succ_s M_s$ 
% and $s\in Ch_{c}(I_c(M_c'))$ where  $M_c' = M_c \cup \{s\}$. 
% \end{enumerate}
% \end{definition} 
\section{Conclusion}
In this paper, we have addressed the challenge of student placements under diversity constraints by introducing the concept of \emph{balanced representation}. Our proposed choice function is distinguished by its satisfaction of maximal diversity, non-wastefulness, balanced representation, and justified envy-freeness.
We demonstrated the practical advantages of our approach through the development of efficient algorithms. 
Future work could explore further refinements to these algorithms and investigate their applicability to other settings where fairness and efficiency are critical.

\section*{Acknowledgments}

This work was partially supported by the Japan Science and Technology Agency under the ERATO Grant Number JPMJER2301.
We thank the anonymous reviewers of AAAI 2025 for their valuable comments.

%Bibliography
\bibliographystyle{plainnat}
\bibliography{balance}

\end{document}